\documentclass[fleqn,11pt]{wlscirep}
\usepackage[utf8]{inputenc}
\usepackage[T1]{fontenc}
\usepackage{cite}
\usepackage{amsmath}
\usepackage{graphicx}
\usepackage{amssymb}
\usepackage{mathtools}
\usepackage{siunitx}
\usepackage{ulem}
\usepackage{soul}
\newcommand{\Ioff}{\ensuremath{I_\mathrm{off}}}
\newcommand{\Ion}{\ensuremath{I_\mathrm{on}}}

\newcommand{\opar}[1]{\ensuremath{\frac{\partial #1}{\partial z}}}
\newcommand{\npar}[1]{\ensuremath{\partial_z #1}}

\newcommand{\red}[1]{ { #1}}

\title{Resolution of the paradox of the diamagnetic effect on the Kibble coil\thanks{Accepted for publication in \href{https://www.nature.com/srep/}{{\it Scientific Reports}}.}} 

\author[1,*]{Shisong Li}
\author[2,+]{Stephan Schlamminger}
\author[2]{Rafael Marangoni}
\author[1]{Qing Wang}
\author[2]{Darine Haddad}
\author[2]{Frank Seifert}
\author[2]{Leon Chao}
\author[2] {David Newell}
\author[3]{Wei Zhao}

\affil[1]{Department of Engineering, Durham University, Durham DH1 3LE, United Kingdom}
\affil[2]{National Institute of Standards and Technology, Gaithersburg 20899, United States}
\affil[3]{Department of Electrical Engineering, Tsinghua University, Beijing 100084, China}

\affil[*]{shisong.li@durham.ac.uk}
\affil[+]{stephan.schlamminger@nist.gov}

\begin{abstract}
Employing very simple electro-mechanical principles known from classical physics, the Kibble balance establishes a very precise and absolute link between quantum electrical standards and macroscopic mass or force measurements. The success of the Kibble balance, in both determining fundamental constants ($h$, $N_A$, $e$) and realizing a quasi-quantum mass in the 2019 newly revised International System of Units, relies on the perfection of Maxwell’s equations and the symmetry they describe between Lorentz’s force and Faraday’s induction, a principle and a symmetry stunningly demonstrated in the weighing and velocity modes of Kibble balances to within {$1\times10^{-8}$}, with nothing but imperfect wires and magnets. However, recent advances in the understanding of the current effect in Kibble balances reveal a troubling paradox. A diamagnetic \red{effect, a force that does not cancel between mass-on and mass-off measurement,} is challenging balance maker’s assumptions of symmetry at levels that are almost two orders of magnitude larger than the reported uncertainties. The diamagnetic \red{effect} , if it exists, shows up in weighing mode without a readily apparent reciprocal effect in the velocity mode, begging questions about systematic errors at the very foundation of the new measurement system.  The hypothetical force is caused by the coil current changing the magnetic ﬁeld, producing an unaccounted force that is systematically modulated with the weighing current. Here we show that this diamagnetic force exists, but the additional force does not change the equivalence between weighing and velocity measurements. We reveal the unexpected way that symmetry is preserved and show that for typical materials and geometries the total relative effect on the measurement is $\approx \SI{1e-9}{}$.
\end{abstract}
\begin{document}
\flushbottom
\maketitle
\thispagestyle{empty}
\clearpage

\section*{Introduction}

Most human activities, especially science, industry, and trade rely on measurements.
The importance of measurement to global society is such that the International System of Units (SI) was created as early as 1875 so that all measurements might be traceable to a single compact set of common standards. For a long historical period the SI standards were formulated by artifacts (man-made or using a property of nature), specific objects preserved in a single location, with limited access. Undeniably inaccessible, the value that such an artifact standard realizes may also vary over time~\cite{stock2015calibration}, introducing dark uncertainties for precision science and high-accuracy engineering~\cite{merkatas2019shades}. Consequently, alternatives to artifact standards have been sought since the beginning of the SI~\cite{flowers2004route}. The first success was measurement by counting events of microscopic particles (e.g., atom, electron, photon, etc), ﬁrst used in time measurements based on atomic clocks, which opened the door for the quantum measurement of things~\cite{mehlstaubler2018atomic,Nakamura889,wcislo2016experimental,katori2011optical}. On May 20, 2019, a new International System of Units, in which all seven base units are deﬁned by physical constants of nature, was formally adopted~\cite{codata17,fischer2016new} and our daily measurement activities have entered into a quantum era. With this quantum revolution of the SI, our measurement system relies now on fundamental constants which are woven into the structure of our universe and are here for all times and for all people, and are no longer tied to physical objects with limited stability and availability. The new SI provides a highly accurate or ultra-sensitive measurement foundation to support explorations that were not possible in the past~\cite{oelker2019demonstration,mcgrew2018atomic}. The change is most profound for mass quantities, where the quest for an atomic or quantum based standard of mass vexed researchers for decades.

To realize the unit of mass at the kilogram level from atomic or quantum standards, two complementing technologies were eventually found, the X-ray crystal density (XRCD) method~\cite{fujii2016realization} and the Kibble balance\cite{Stephan16}.The XRCD method relies on the mass of the electron, which is given by the Rydberg constant and defined fundamental constants. Using mass spectroscopy and scaling that takes advantage of a nearly perfect single crystal silicon sphere, the electron's mass can be scaled thirty orders of magnitude to the kilogram level with a relative uncertainty of \SI{1e-8}{}.~\cite{bartl2017new} The realization of the kilogram via the Kibble balance relies on the perfect symmetry of Maxwell’s equations and can reach a similar uncertainty~\cite{NRC,NIST}, thanks to some Nobel prize winning quantum physics~\cite{zimmerman2010quantum,haddad2016bridging}.

In the 1980s, the discovery of the quantum Hall effect by Prof. von Klitzing \cite{klitzing1980new} provided a catalyzing piece in the quest of a quantum mass standard. It was almost immediately recognized that the quantized resistance standard \cite{jeckelmann2001quantum} that resulted from von Klitzing's work could be combined with the Josephson effect that had been theoretically postulated in 1962 \cite{1962Jose} and experimentally verified a year later \cite{anderson1963probable} allowing the measurement of electrical power solely based on quantum effects. Once electrical power could be measured via quantum standards, a machine that precisely compares electrical to mechanical power would allow the quantum realization of mechanical power given by force times velocity. Velocity is easily measured as a unitless fraction of speed of light and the force could be, for example, the weight of a mass standard in the gravitational ﬁeld of the Earth. All that is needed is a precise tool that can compare mechanical to electrical power.

Luckily, such a tool, a comparator, existed. It was proposed in 1976 by Dr. Bryan Kibble~\cite{Kibble1976}, a metrologist at the National Physical Laboratory in the United Kingdom. Kibble's invention was initially named a watt balance, emphasizing that it compares mechanical to electrical power, since the watt is the unit of power, both electrical and mechanical. Kibble passed away in 2016, and the watt balance was renamed Kibble balance to honor his contributions to metrology. The core of Kibble's idea lies in a symmetry of electromagnetism, described by Maxwell's equations~\cite{schlamminger2017quantifying}. In a nutshell, it can be described as follows: The energy of a current-carrying loop (a coil with one turn) in a magnetic field is given by the product of current, $I$ and the magnetic flux, $\phi$ threading the coil. The Lorentz force in the vertical direction $F_z$ on the coil is the negative derivative of the energy of this loop with respect to its vertical position, $z$.
\begin{equation}
F_z = -\npar{\phi} I
\end{equation}

In this text, we use the abbreviation $\npar{A}:=\opar{A}$ for the partial derivative of a quantity $A$ with respect to $z$. The current is easy to measure, but not the derivative of the flux through the loop. Here is where the symmetry of nature comes to the rescue: Moving the coil in the magnetic field produces an induced electro-motive force between both ends of the coil. By Faraday's law of induction, the induced voltage, $U$ is proportional to the product of the derivative of the flux times the vertical velocity, $v_z$, of the wire loop.
\begin{equation}
U = -\npar{\phi} v_z
\end{equation}
Both equations can be combined to obtain the watt equation that shows the equivalency of mechanical to electrical power, and conveniently the hard to measure flux derivative vanishes.
\begin{equation}
F_z v_z = U I \label{eq:Fv_UI}
\end{equation}
By using the weight of a mass $F_z=mg$ for the force and the quantum measurement of the electrical power $UI=C f^2 h$, where $f$ is the frequency that is used to drive the programmable Josephson junction voltage array and $C$ is a known constant that depends, for example, on how many Josephson junctions are used, the mass can be written as
\begin{equation}
m = \frac{Cf^2h}{g{v_z}}. \label{eq:Kibblemass}
\end{equation}

\begin{figure}[tp!]
\centering
\includegraphics[width=1\textwidth]{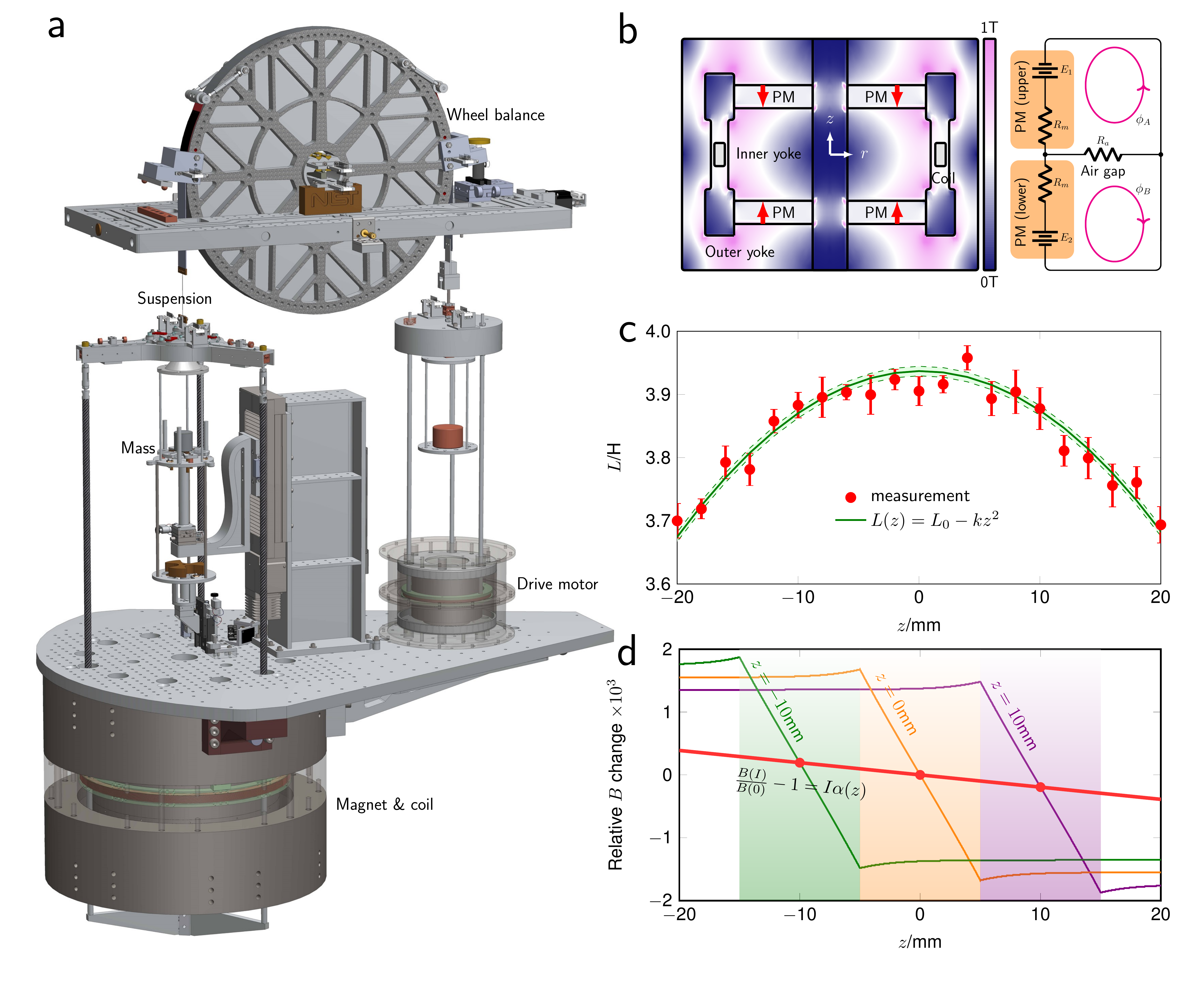}
\caption{The magnet system in a Kibble balance and the coil-current effect. (a) presents the major elements in the fourth generation Kibble balance experiment at NIST. The left subplot of (b) is the sectional view of a typical permanent magnet system with symmetry, where the color map denotes the $B$ field distribution. The right subplot presents an equivalent electrical circuit of the air-gap type magnet system, where $R_m$ is the magnetic reluctance of the permanent magnet, $R_a$ the magnetic reluctance of the air gap, $E_1$, $E_2$ respectively the magnetomotive force of upper and lower magnets. (c) shows a typical measurement of the coil inductance (frequency extrapolated to DC) as a function of coil vertical position $z$. With a up-down symmetrical magnet, it can be written as $L=L_0-kz^2$. (d) shows the relative magnetic field change due to the coil current in such magnet systems. The plot shows the magnetic field with a plus current \Ioff, which produces 4.9\,N magnetic force. The red curve is an average magnetic field for the coil, and this field slope has been verified at BIPM as $\frac{B(I)}{B(0)}-1=I\alpha(z)$, where $\alpha$ is a linear function of $z$. Note that the field distribution with \Ion\ is an image of \Ioff\
symmetrical to $B(0)$. }
\label{fig1}
\end{figure}

Figure \ref{fig1}(a) shows a typical Kibble balance. Two large components are apparent: the magnet and the wheel. The wheel is a particular choice for a part that can be used as a moving and weighing mechanism. The wheel allows the comparison of electromagnetic force and mass weight while also providing the coil's motion needed for the velocity mode. Up to  the 2000s, several different types of magnet systems were used \cite{npl1,nist3,metas1,nimx,nist1}. Over time, the field matured, and the magnet systems' design converged to what is known as the air-gap type, yoke-based magnetic circuits \cite{BIPM,NISTmag,LNE,METAS,KRISS,NIM,MSL,UME}. Figure \ref{fig1}(b) shows a typical construction of such a permanent magnet system. The permanent magnetic circuit's significant advantage is that it can supply a strong (several tenths of a tesla), uniform, and robust magnetic field without an active energy source.

While the description using the derivative of the flux is accurate and was used initially by Kibble, these days, the researchers use a different description of the same numerical quantity, the so-called geometric factor, or flux integral. The geometric factor is obtained by integrating the horizontal component of the magnetic flux density $B$ that is perpendicular to the wire  with a length $l$ that forms the coil. It is abbreviated as $ Bl $, and by virtue of Green's theorem it is the same as $\npar{\phi}$. For the rest of the article, we consider the possible causes and consequences of inevitable imperfections in the symmetry, so that there are two different geometric factors, one for weighing mode, $(Bl)_w$, and one for velocity mode, $(Bl)_v$.
\red{A succinct equation for the relationship of both geometric factors was suggested by  Robinson~\cite{NPL}. The widely accepted equation is}
\begin{equation}
(Bl)_w=(Bl)_v(1+\alpha I+\beta I^2), \label{eq:BlW}
\end{equation}
where $I$ is the current circulating in the coil during weighing mode. Observe that the current dependence of $(Bl)_w$ is canceled to first order through a common reversal trick in the design of the weighing mode: Two measurements must be made, one without and one with mass on the mass pan of the balance. The balance, however, can be biased with a tare weight, $m_t\approx m/2$, such that the currents in the coil have the same absolute value but opposite signs. The forces on the balance for the two states are
\begin{eqnarray}
\begin{cases}
\mbox{mass-on}:\Ion (Bl)_v(1+\alpha \Ion+\beta \Ion^2) +mg  = m_t g ~~\label{eq:massOn}\\[1ex]
\mbox{mass-off}:\Ioff (Bl)_v(1+\alpha \Ioff+\beta \Ioff^2) = m_t g.\label{eq:massOff}
\end{cases}
\end{eqnarray}
The tare weight is adjusted such that the currents are symmetric, $\Ion=-\Ioff$, and it is sufficient to work with the variable $I:=\Ioff$. By subtracting the mass-off equation from the mass-on equation in (\ref{eq:massOn}), the mass can be obtained as
\begin{equation}
m =\frac{(Bl)_v}{g} \left(  2 I +  2 \beta I^3 \right),
\end{equation}
where as mentioned before $(Bl)_v$ is obtained from the velocity mode. By using symmetric currents, all terms containing  $\alpha$ vanish. The only remaining systematic term, $2\beta I^3$ is very small, $2\beta I^3/(2I)\approx 10^{-9}$.~ \cite{linonlinear,linonlinear2} Although the term is small, it is measurable by using different mass values on the Kibble balance, e.g., $m/2$, $m$, $2m$. This process is possible in the new SI, because multiple and sub-multiples of masses can be generated without having to resort to Kibble balances using a classical scheme to subdivide masses~\cite{NIST,NRC}. In summary, the Kibble principle is preserved when symmetric currents are applied during weighing mode, because the dominant term of the dependence of the magnetic field on the weighing current drops out. The next to leading order effect is small and can be compensated for using ancillary measurements. For the remainder of this manuscript, we assume $\beta=0$ without altering the main conclusion, but simplifying the equations.

We next take up a question that has vexed the Kibble balance community for years. What if extraneous magnetic forces act on the weakly magnetic materials of the coil? Put another way, what if the coil is a magnet? We use the term weakly magnetic materials for materials that exhibit diamagnetic or paramagnetic behaviour, in other words material whose magnetization, $M=\chi H$ depends linearly on the applied external field $H$. The proportionality factor is given by the volume susceptibility $\chi$ and is negative for diamagnetic and positive for paramagnetic materials. It is impossible to build a coil without using weakly magnetic materials. The magnet wire used to wind the coil is made from copper which is diamagnetic with $\chi\approx- 10^{-5}$. {The diamagnetic force has been impressively demonstrated by levitating a diamagnetic object, e.g. graphite~\cite{waldron1966diamagnetic}, organics~\cite{beaugnon1991levitation}, water~\cite{ikezoe1998making}, living cell~\cite{winkleman2004magnetic}, even a frog~\cite{frog}, in a magnetic field.} The force on a very small element with volume $V$ of weakly magnetic material in the air gap of a permanent magnet is given by
\begin{equation}
F_\chi =\frac{\chi V}{\mu_0}B \npar{B}.
\end{equation}
There is a constant static force acting on the coil, but it is common to the mass-on and mass-off measurement, similar to the coil's weight, and it will drop out in the difference of the mass-on and mass-off measurement. \red{To be clear, the diamagnetic force on the coil is well-known but is thought to drop out in the reversal of the current in force mode. The systematic described below is named the diamagnetic effect in force mode. It is the diamagnetic force that does not cancel between the two measurements in force mode.}

Nevertheless, a systematic bias,\red{the diamagnetic effect},   cannot be ruled out, because $B$ is not a constant, rather it is a function of current in the coil according to equation~(\ref{eq:BlW}). Consequently, $F_\chi$ must be a function of current also. The difference between mass on and off would be
\begin{equation}
\Delta F_\chi = \frac{2 \chi V}{\mu_0}I \npar{\left( B_v^2 \alpha \right)} \label{eq:Delta_F_chi}
\end{equation}
so that the quadratic nonlinearity no longer cancels and we are forced to consider $\alpha$. Up until 2017, $\alpha$  was assumed  to be constant, independent of the coil position, and dispensed with. A notable article published that year associated $\alpha$ with the reluctance effect \cite{li17}. The reluctance effect can be explained by considering the magnetic energy stored in the magnetic field surrounding the coil due to the constant current during weighing, $E=1/2\,LI^2$ where $L$ is the self inductance of the coil in its surroundings. Once again, a force arises as a result and in the direction of any gradient of the magnetic energy. The vertical component of this force  can be written as $F_z=-1/2\,I^2\npar{L}$. The force points toward the maximum of the inductance, usually at the middle of the symmetry plane of the coil magnet system. This principle is well known from solenoid actuators, where an iron slug is retracted into a solenoid when it is energized with current. Here, the slug (the magnet and yoke material) is fixed, while the coil is free to move in the $z$ direction. The inductance $L(z)$ depends mostly on the symmetry of the shape and magnetic properties of the yoke and not on the permanent magnet material. For an ideal yoke $ L =L_0-kz^2$ is a quadratic function of $z$ with $z=0$ in the symmetry plane of the yoke, see figure~\ref{fig1}(c). Interestingly, the reluctance force can be interpreted as a force produced by an additional magnetic field, so instead $F_z=-1/2\,I^2\npar{L} = (Bl)_\mathrm{add} I $, and hence $(Bl)_\mathrm{add}=-1/2\,I\npar{L}$. As described in figure \ref{fig1}(d), experiments at the BIPM prove that this additional magnet field does, in fact,  exist\cite{BIPMmag2017,li17}. Hence, the parameter $\alpha$ introduced in equation~(\ref{eq:BlW}) can be written as
$\alpha = -k z/(Bl)_v$.

The partial derivative in equation~(\ref{eq:Delta_F_chi}) can be rewritten as $\npar{B_v^2\alpha} = \alpha \npar{(B_v^2)} +B_v^2 \npar{\alpha}$. The magnet systems for the Kibble balances are often designed such that $\npar{B_v}=0$ rendering the first term insignificant. The second term evaluates to $-B_v^2 k/(Bl)_v $, and the relative size of the effect can be obtained from equation~(\ref{eq:Delta_F_chi}) as
\begin{equation}
\frac{\Delta F_\chi}{mg} =  -\frac{\chi}{\mu_0}\frac{A_c}{Nl}  k. \label{eq:diaF}
\end{equation}
Here, we are formulating the effect on the wire while considering multiple turns, so the volume of the wire, $V$ has been replaced by the product of the wire cross sectional area $A_c$, the length $l$ and the number of turns $N$.
The derivation will also work for non-current carrying elements, like the coil former or structures mounted on the coil, but the equations are more insightful for the wire. The relative effect consists of three  factors and typical values are  $\chi/\mu_0 =\SI{-8}{\meter \per \henry}$, $A_c/(Nl)=\SI{200}{\milli\meter^2}/(1057\cdot  \SI{834}{m})=\SI{2.27e-10}{m}$, and $k=$\SI{550}{\henry \per \meter^2}. Multiplying the three factors together yields a relative force of \SI{1e-6}{}. An amount that is more than 100 times larger than the combined relative uncertainty reported by the best experiment in the world.

Here we reach an impasse. The paradox. On the one hand, the above summary of current reasoning, modeling and experimentation supports the conclusion that the \red{diamagnetic effect in force mode does} exist. On the other hand, measurements of the Planck constant using two completely different methodologies (XRCD and Kibble balance) agree to within \SI{1e-8}{}, supporting the conclusion that it doesn't. Where is the truth?

A possibility that must be considered is that there is a common bias, or intellectual phase-lock among the experiments. After all, the highest precision Kibble balances share similar design parameters, and the community was driven by a common goal to seek a consensus value. Perhaps the relative size of the effect does not vary much from balance to balance. Being common mode to all, it would not be observed. But values of the Planck constant were compared among all Kibble balance {\it and} XRCD methodologies. To support such a bias among the balances requires intellectual phase lock across the competing methods and multiple laboratories on a global scale. This seems highly unlikely in a metrology community fiercely committed to objectivity.

Another possibility that must be considered is that the \red{diamagnetic effect in force mode} doesn't exist.  The deniers of this effect likened the \red{force produced by} it to the fictional force that Baron Munchausen used to  pull himself out of a mire by his own hair -- clearly in violation of Newton's third law. They argue, that the current in the coil cannot exert \red{an additional and current dependent} force on itself. This force, however, is between the magnet system, altered by the current, and the coil, similar to the reluctance force that undoubtedly exists (A detailed analysis can be found in the Supplementary Information). Given the state of knowledge, it seems logical to suggest an experiment be performed to measure the \red{effect} directly. Unfortunately, this is exceedingly difficult. According to equation~(\ref{eq:diaF}) the \red{effect} depends only on  variables that are, for the most part, impossible to modify for a given Kibble balance. These are instruments designed to maintain absolutely constant physical, magnetic, and electrical geometries save for one coordinate. Changing the mass, and hence the current in the coil, will not change the relative contribution of the diamagnetic force. The only variable sometimes available is the coil geometry $A_c/(Nl)$, but even that is not simple. Several Kibble balances have multiple coils wound on a single former, and the Kibble experiment can be performed with different coils or different coil combinations. Unfortunately, the relative contribution of the diamagnetic effect does not change as long as all coils are immersed in magnetic flux produced by the same magnetic system, regardless if they are active (used in the experiment) or not. In summary, it is conceivable that a relative bias as large as \SI{1e-6} exists in all Kibble balance experiments.

In this article, we will solve the paradox of the diamagnetic \red{effect in force mode}. The surprising result is that the diamagnetic \red{effect} exists, but we find a symmetric effect in the velocity mode. By combining the measurements taken in velocity mode with those made in weighing mode, the bias introduced by the diamagnetic \red{effect} is canceled. These counteracting biases explain the paradox, restore confidence in the foundation of the new SI mass, and have never been described in the literature. The result is simple and satisfying: the symmetry of the Kibble balance experiment once again self corrects, and the diamagnetic effect vanishes in the combined result. This new finding will relax the requirements on the materials that the coil and components attached to it are made from. Weakly magnetic materials can be used in these cases. Still, one has to be careful not to use ferromagnetic materials, because materials with a nonlinear response to the external field are not covered by this symmetry.

\section*{Results}
\subsection*{Analytical result of the diamagnetic effect in velocity measurement}

In the previous section, we have argued that the diamagnetic \red{effect}  exists and that it produces a large relative bias in the weighing mode of Kibble balances. The bias is so large that Kibble balances would not be able to make precise measurements. Here we show that the bias in the weighing mode is cancelled by an identical bias in the velocity mode and the Kibble principle holds.

We start by rewriting the self inductance of the coil $L(z)$ with $N$ turns according to the derivation in the methods section.  In a cylindrical air-gap with a mean radius  $r_a$, a radial width of $w_a$,  and a height $2h_a$, the inductance is given by
	\begin{equation}
	L(z)=L_0-\pi \mu_0 N^2\frac{ r_a}{w_a} \frac{z^2}{h_a} \;\;\Longrightarrow\;\; k=-\frac{1}{2}\frac{\partial^2 L}{\partial z^2} = 2\pi \mu_0 N^2 \frac{r_a}{A_a},
	\label{inductance}
	\end{equation}
where $A_a=2w_ah_a$ denotes the cross-sectional area of the air gap. By employing a cross sectional area for the coil, the relative size of the diamagnetic \red{effect}  can be written compactly as
\begin{equation}
\frac{\Delta F_\chi}{mg} =  -\frac{\chi}{\mu_0}\frac{A_c}{Nl}  \frac{2\pi r_a \mu_0N^2}{A_a}=-\chi\frac{A_c}{A_a}\frac{r_a}{r_c}. \label{eq:diaF2}
\end{equation}

\begin{figure}[tp!]
		\centering
		\includegraphics[width=0.8\textwidth]{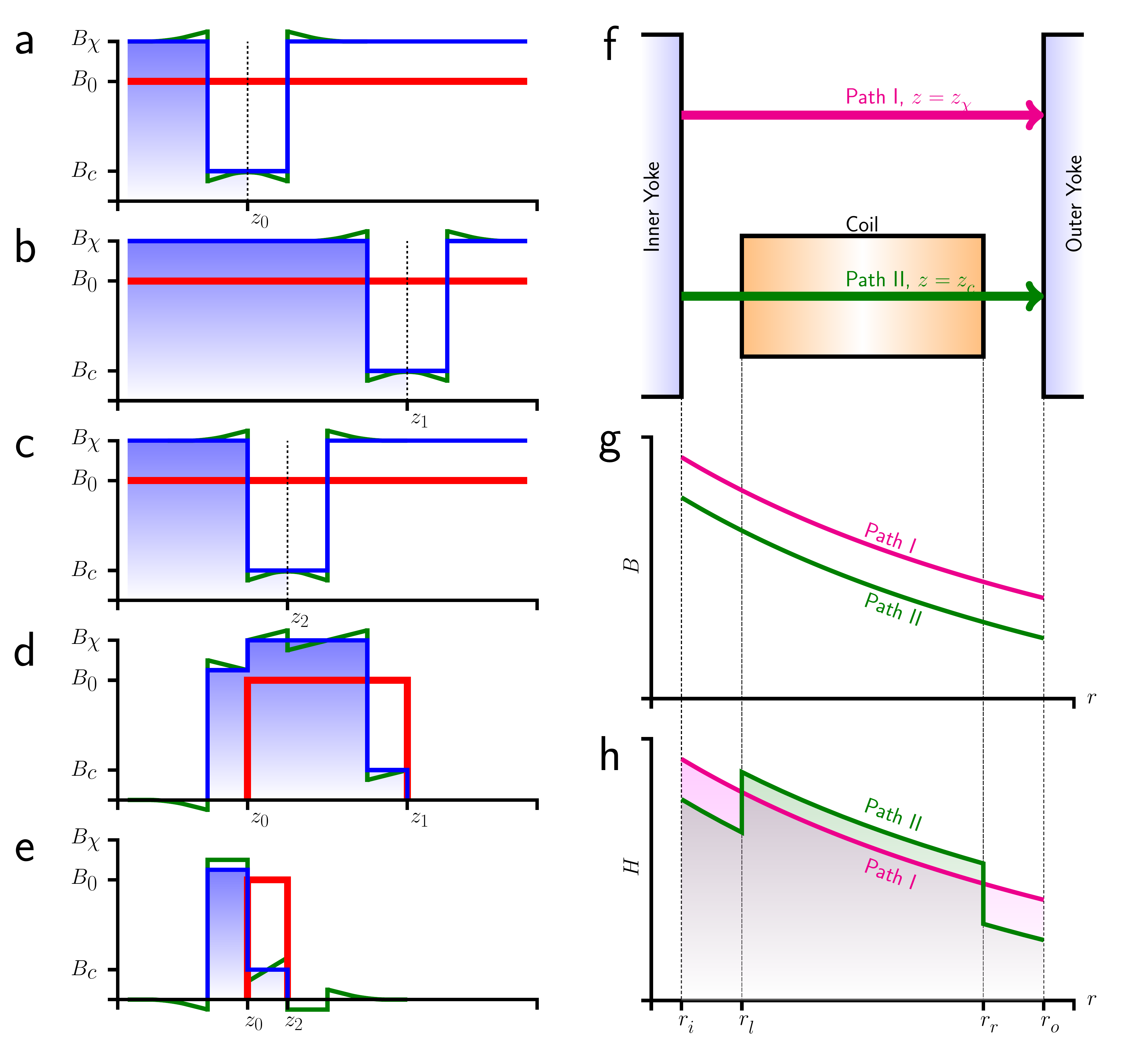}
		\caption{{A qualitative illustration of the} magnetic field distribution at different coil positions. (a)-(c) show the $B(r_c, z)$ curves at three different vertical position $z_0$, $z_1$ and $z_2$. The red curves are the magnetic profile when the coil susceptibility is zero. The blue curves are the first order approximation of $B$ field curve with a diamagnetic coil, $\chi<0$. The green curves are profiles of the diamagnetic coil with higher order approaching. (d) and (e) present the $B$ field difference under two configurations: $(z_1-z_0)>2h_c$ and $(z_2-z_0)<2h_c$ (coil region overlap). (f) show two paths horizontally across the air gap, respectively at $z_\chi$ and $z_c$. (g) and (h) present the $B$ field and the $H$ field distributions along two paths. }
		\label{fig2}
\end{figure}

Next, we investigate what happens when a diamagnetic material is introduced to the air gap. The left plots in figure~\ref{fig2} show the magnetic flux density as a function of vertical position. Before the material is introduced, the flux density is constant throughout the gap (red line). A constant flux density for the air gap is assumed to keep the explanation simple, but is not necessary for the theory to work. Adding the coil, here with $\chi<0$, changes the flux profile. A perfectly nonmagnetic coil would have no effect, but the vertical section occupied by the coil now restricts the flux, due to the increased magnetic reluctance of the diamagnetic material in that part of the gap. The total flux produced by the permanent magnet redistributes itself, and, as a result, the flux density in the empty space increases in direct proportion to the reduction of flux through the space occupied by the coil.

For $\chi<0$, compared to the situation without the coil, $B_0$, the value of the flux density is lower at the coil ($B_c$) and higher in the rest of the gap ($B_\chi$). In the physical system, there are nonlinear effects near the edges, shown by the green curves in figure~\ref{fig2}. Again, these are not important for the simplified explanation of the effect and can be ignored.

The magnetic flux threading through the coil can be obtained as the integral from the bottom of the air gap to the middle of the coil, indicated by the blue shaded region for the coil in three different vertical positions in figure \ref{fig2} (a) through (c).

As mentioned above, the induced voltage in velocity mode is proportional to the derivative of the magnetic flux through the coil with respect to time. The flux for the baseline position of the coil $z_0$ is shown in (a), while the flux for the positions $z_1$ and $z_2$ are shown in (b) and (c). The difference in flux with respect to the coil at baseline for these positions is depicted in (d) and (e), respectively. We assume that the coil moves through the gap along a fixed trajectory with the same constant velocity in the $z$ direction for the case when coil susceptibility is zero, and then again when it is $\chi$. The relative difference of the flux density change between these scenarios, and, hence, the induced voltage is given by
\begin{equation}
    \frac{\Delta U_\chi}{U}=\frac{B_\chi}{B_0}-1. \label{eq:Delta_U}
\end{equation}

The flux density $B_\chi$ can be calculated assuming that the total magnetic flux through the air gap remains the same.
At $r=r_c$, the flux integration vertically through the whole air gap can be written as
\begin{equation}
    2h_aB_0=2h_cB_c+(2h_a-2h_c)B_\chi\; \Longrightarrow\; \frac{ \left(B_\chi-B_0\right)}{\left(B_c-B_0\right)} = - \frac{2h_c}{2h_a-2h_c},
    \label{eq:boundary1}
\end{equation}
where the negative sign indicates that $B_\chi>B_0$ and $B_c<B_0$ when diamagnetic material is introduced. For paramagnetic material, $B_\chi<B_0$ and $B_c>B_0$.

The ratio of the change from $B_0$ of $B_\chi$ and $B_c$ to $B_0$ is identical to the ratio of the height of the occupied gap to the height of the empty air-gap, since $2h_a$ and $2h_c$ denote the height of the air gap and the coil, respectively.

In an actual magnet system, the magnetic height of the air gap $2 h_a$ differs from the geometrical height of the air gap $2 h_\mathrm{geo}$ as one would measure with a ruler. Due to fringe fields, $h_a>h_\mathrm{geo}$. We assume the magnetic height of the air gap is known.

For the typical large permeabilities  of the yoke materials, the metal on each side of the air gap is a magnetic equipotential surface. Hence, the magneto motive force  over the air gap given by $\int_{r_i}^{r_o}H(r)\mbox{d}r$ with $r_i$ and $r_o$ denoting the inner and outer radius of the air gap, does not change when the coil is introduced and is independent of the vertical position $z$ where the integration is performed. The magnetic field $H$ is the magnetic flux divided by the permeability, $H=B/(\mu_0(1+\chi))$ and is a function of radius and height $H(r,z)=H(z)r_c/r$, where we have used the fact that the field drops off as $1/r$ and $H(z)$ is the field at the mean coil radius $r_c$.

Two paths of integration through the coil  at $z=z_c$ and  in the empty air gap ($z=z_\chi$) are shown in~\ref{fig2} (f). Integration along these paths  yield
\begin{eqnarray}
    \int_{r_i}^{r_o}\frac{H(z_\chi) r_c}{r}\mbox{d}r &=& \int_{r_i}^{r_o}H(r,z_c)\mbox{d}r \nonumber\\[2ex]
    \int_{r_i}^{r_o}\frac{B_\chi r_c}{\mu_0r}\mbox{d}r&=&\int_{r_i}^{r_l}\frac{B_cr_c}{\mu_0r}\mbox{d}r+\int_{r_l}^{r_r}\frac{B_c r_c}{\mu_0r(1+\chi)}\mbox{d}r+\int_{r_r}^{r_o}\frac{B_c r_c}{\mu_0r}\mbox{d}r\nonumber\\[2ex]
    \int_{r_i}^{r_o}\frac{B_\chi}{r}\mbox{d}r&\approx&\int_{r_i}^{r_o}\frac{B_c}{r}\mbox{d}r-\int_{r_l}^{r_r}\frac{\chi B_c }{r}\mbox{d}r,
    \label{eq:boundary2}
\end{eqnarray}
where $r_l=r_c-w_c/2$ and $r_r=r_c+w_c/2$ denote the inner(left) and outer(right) edge of the coil. The quantities $B(r)$ and $H(r)$ as a function of $r$ for both integration paths are shown in \ref{fig2} (g) and (h), respectively. Integrating the terms, approximating the resulting logarithms in a Taylor series of first order and combining the result with equation~(\ref{eq:boundary1}) gives
\begin{equation}
    B_0=B_\chi \left( 1+\chi\frac{w_c h_c}{w_a h_a}\frac{r_a}{r_c} \right)
\end{equation}

With equation~(\ref{eq:Delta_U}) the relative change that the magnetic material has in the velocity mode can be stated as
\begin{equation}
    \frac{\Delta U_\chi}{U}=\frac{B_\chi}{B_0}-1\approx-\chi\frac{A_c}{A_a}\frac{r_a}{r_c}. \label{eq:diaV}
\end{equation}
As before, $A_c=2h_cr_c$ and $A_a=2h_ar_a$ denote the cross-sectional areas of the coil and the air gap, respectively.

Equation (\ref{eq:diaV}) shows the relative change of the induced voltage in the velocity mode is identical to the relative change in force mode, see equation (\ref{eq:diaF2}). The robustness of Kibble's reciprocity to deviations from the ideal experimental setup without magnetic materials, are caused by a strong symmetry in the underlying physics. Without that robustness the Kibble balance would not be the success that it has been in metrology. The relative differences of the measured force and in voltage from the corresponding ideal theoretical values in the absence  of weakly magnetic materials are  given by
\begin{equation}
    \frac{U_{\mathrm{real}} - U_{\mathrm{ideal}}}{U_{\mathrm{ideal}}} =\frac{F_{\mathrm{real}} - F_{\mathrm{ideal}}}{F_{\mathrm{ideal}}} \approx -\chi\frac{A_c}{A_a}\frac{r_a}{r_c}. \label{eq:result}
\end{equation}
Since $r_a \approx r_c$, the relative effect is proportional to the magnetic susceptibility and the cross-sectional filling ratio of the air gap. The latter denotes how much of the cross-sectional area of the air gap is taken up by the coil. With typical values,  $\chi=-10^{-5}$ and $A_c/A_a=0.1$, the relative difference between the real and ideal numbers is \SI{1e-6}{}. In conclusion, the diamagnetic force with a relative magnitude of \SI{1e-6} about 100 times larger than the reported relative uncertainties exists. But the results of the Kibble balance experiments are not affected by it, because the same relative bias will be introduced in the velocity mode. In the combination of the measurement results from force and velocity mode, the effect cancels perfectly.

The derivation above has been made using ideal geometries to show the powerful and simple idea. But, the theory holds for more complex and realistic field situations, as is discussed in the ‘Methods’ section.

\subsection*{Numerical verification}

Numerical verification of a relative force change that is as small as \SI{e-6} is impossible.
{  Since engineering tasks are rarely concerned with effects that small in size, commercial finite element programs are not optimized for the precise prediction of these small effects. At this order of magnitude their results cannot be trusted. To overcome their limitations and to be able to use commercial finite element analysis (FEA) software, we invented a new technique that we name differential FEA (dFEA). While more information on dFEA can be found in the supplemental information, the following paragraphs explain the general idea.} \red{All} effects discussed here are proportional to the magnetic susceptibility $\chi$ and it can be used as a parameter to verify the result. Setting  $\chi$ to a large value amplifies the relative change in force and voltage. With $\chi\approx$\SI{1e-2},  relative effects of \SI{1e3} are achieved. Although the theory is only weakly dependent on geometry and independent of the size of the magnetic field,  typical values are used. A magnetic flux density of $B_0=0.54\,$T was chosen, and it requires $+/-11.6$\,A in a single turn to produce half the weight of a kg standard in positive/negative vertical direction. \red{Ansoft} a commercial \red{FEA} software was used to calculate the force produced on a coil consisting of a single turn. \red{For all calculations shown below an adaptive mesh strategy and a nonlinear solver were used.} \red{ The calculations were performed with} five different values  for the magnetic susceptibility of the coil wire  ranging from, -0.01 to 0.01. The precise result of the calculation depends on how the geometry is meshed by the FEA software. To avoid any bias in this investigation, the mesh is only calculated once and fixed for all subsequent calculations. For each $\chi$ value,  the force on the coil without current is calculated. Then the forces for positive and negative currents are calculated. From both the null result is subtracted. This differential approach suppresses systematic errors in the calculation due to meshing and rounding of the small effects.

\begin{figure}[tp!]
		\centering
		\includegraphics[width=1\textwidth]{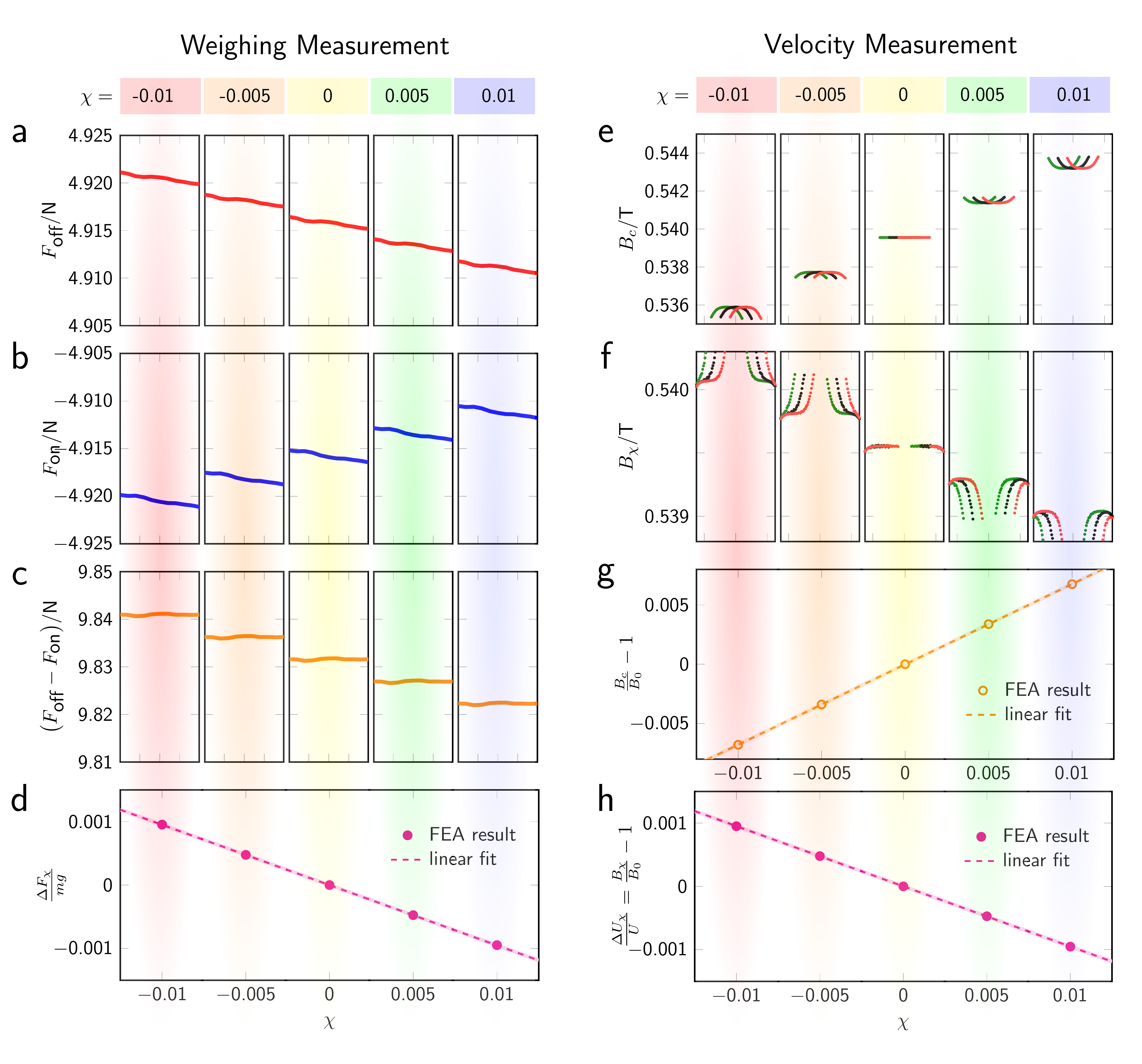}
		\caption{Results of the magnetization effect in weighing and velocity measurements. (a)-(c) Results of magnetic force as a function of coil position -10\,mm\,$\le z\le$\,10\,mm for five different magnetic susceptibilities. The current in the coil was equivalent to one turn with \Ioff=-\Ion=11.6\,A. (d) The relative change in force difference as a function of $\chi$. Note that $mg$ is defined at force difference at $\chi=0$.
		(e) The $B_c$ distribution at $z=\pm6$\,mm and $z=0$\,mm with different $\chi$ values. No current is assigned in the calculation. (f) shows the $B_\chi$ field distribution. (g) presents the relative magnetic field change of $B_{c}$ and (h) shows the change of $B_{\chi}$. 
		}
		\label{fig3}
	\end{figure}
	
The calculated force differences with positive (mass-off) and negative (mass-on) currents are shown in figure \ref{fig3}(a) and (b), respectively. In each subplot, the force is given as a function of coil position $z$ with $\SI{-10}{\milli\meter}\,\le z\le\,\SI{10}{\milli\meter}$ for the five $\chi$ values. The clearly visible slope in each subplot, is caused by the reluctance force, in  agreement with the theoretical model discussed in~\citen{li18}. The slope is the same  for both current directions and independent of $\chi$. Hence, in  the force difference, shown in figure \ref{fig3}(c), the slope vanishes and the difference is nearly independent $z$, exactly as described in \citen{li18}. Neglecting the slopes, the observed values of $F(I_{\mbox{off}})$ and $F(I_{\mbox{on}})$ at a given point, for example $z=0$, change with $\chi$. The change of the force difference relative to the weight of a 1 kg mass at $z=0$ is shown in figure \ref{fig3}(d).

The size of the force (difference) depends on $\chi$. Diamagnetic materials ($\chi<0$) yield larger absolute values for the forces for both current directions. Unlike the slopes, this effect does not cancel by subtracting mass-on from the mass-off measurement. The effect is clearly visible in figure \ref{fig3}(c) where the force differences are plotted.

A linear dependence of the force differences on $\chi$ is observed, see the dashed line in figure \ref{fig3}(d). The slope of the line can be obtained by a numerical regression to the calculation results, and by using the regression coefficients the effect can be scaled down to small $\chi$ values whose results would otherwise be in the rounding error of the numerical analysis. For $\chi=-1\times10^{-5}$ a relative change of \SI{9.50e-7} is obtained, in very good agreement to the theoretically obtained result of $1\times10^{-6}$. The numerical results confirm the theoretical analysis, as well as the existence of the diamagnetic force.

The same FEA calculation can be used to estimate the effect in the velocity mode. Here, we calculate the magnetic flux density in the air gap for a coil that does not carry any current for the five susceptibilities discussed above. As in the text above, two symbols are used to describe the flux density in the air gap in the presence of magnetic material. At regions that the coil occupies we use $B_c$ and all other regions $B_\chi$.

Figure \ref{fig3}(e) and (f) show the magnetic flux densities $B_c$ and $B_\chi$ as a function of $z$, with the coil at three different positions $z_c$, $z_c=\SI{-6}{\milli\meter}$ in green, $z_c=\SI{0}{\milli\meter}$ in black, and $z_c=\SI{6}{\milli\meter}$ in red. For a given $\chi$, one of the two quantities $B_c$ and $B_\chi$ is larger and the other smaller than $B_0$, which can be seen in the two middle panels with $\chi=0$. The curves of $B_\chi$ show a transient step at the border close to $B_c$. This is an artifact of the FEA calculation which cannot reproduce the perfect step function in $B$ that would be present at the boundary in the real world, see figure~\ref{fig2} (b). We believe that the transient has no influence on the conclusion, especially since its integral evaluates to zero. For any $\chi\ne0$,  $B_\chi-B_0$ is about a  seventh of $B_0-B_c$, and, hence, according to equation~(\ref{eq:boundary1}), the effective gap height is about eight times the coil height, such that  $h_c/(h_a-h_c)=1/7$.

To consolidate this assertion, a summary of the relative difference of $B_c$ and $B_\chi$ with respect $B_0$ are shown in figure~\ref{fig3}(g) and (h). Both figures are plotted for $z_c=0$. The former shows $B_c(0)$ the latter $B_\chi$ near the end of the gap, both are relative to $B_0$ at the same locations. Similar to the results in the force mode, the results are linear with respect to the chosen magnetic susceptibility and a regression to the calculation results is performed. From the regression coefficients, $B_{\chi}/B_0-1$ can be calculated for small $\chi$, a result that would be unobtainable directly from finite element analysis. For $\chi =$\SI{-1e-5}, $B_{\chi}/B_0-1$ is \SI{9.51e-7}{}. For comparison, the relative effect in force mode for the same $\chi$ was \SI{9.50e-7}{}. The calculated relative effects in force mode and velocity mode agree remarkably well (the difference of \SI{1e-9} is negligible compared to the numerical uncertainty of the FEA calculation).

The summary of this section is given in the last row of figure~\ref{fig3}. The left graph shows the relative bias that is incurred in force mode as a function  of the magnetic susceptibility of a weakly magnetic coil. The right graph shows the relative bias incurred in velocity mode as a function of the same $\chi$. The results are identical, the relative biases depend linearly on $\chi$. For the model discussed here the slope of the line is approximately $-1/10$, which corresponds to the fraction of the cross sectional area of the air gap that is filled by the coil. So, a Kibble balance with this geometry and a weakly magnetic coil would produce  values for both modes that differ by $-\chi/10$ compared to the same balance that has a completely nonmagnetic coil. However, when the results from the force and velocity mode are combined according to equation~(\ref{eq:Fv_UI}), the relative biases cancel each other and the mass measured by the Kibble balance with a weakly magnetic coil is identical to the mass measured by a Kibble with  a non-magnetic coil.

\section*{Discussion}

The work that led to this article accomplished four tasks.
\begin{enumerate}
\item We have shown that the diamagnetic \red{effect in force mode }exists and its relative magnitude can be as large as \SI{1e-6}{}.
\item We have discovered a corresponding effect in velocity mode that completely cancels out the effect of the diamagnetic force in the Kibble balance experiment.
Such an effect has never been described before in the literature.
\item We have developed a new technique to calculate very small magnetic effects caused by weakly magnetic materials using finite element analysis.
\item By using the newly developed technique we could verify the existence of (1) and (2) and show that they have the same relative size within the numerical uncertainty.
\end{enumerate}
Below we summarize the most important points for these accomplishments.

The  force described by the diamagnetic \red{effect}  exists and it is large ($\approx$\SI{1e-6}) compared to the relative uncertainties that Kibble balances report ($\approx$\SI{1e-8}). A Kibble balance requires a coil immersed in a magnetic field. Often the magnet wire is made from copper that is weakly diamagnetic with $\chi=-$\SI{1e-5}{}. Without current a diamagnetic force on the coil wire exists, but it is a constant force comparable to the weight of the coil and will not impact the result. What is understood as the diamagnetic \red{effect}  is caused by the current in the coil during the weighing measurement. This  current generates an additional magnetic field which interacts with the magnet system in what is known as back-action.
Due to the back-action, the diamagnetic force is no longer constant, but proportional to the current in the coil, and, hence, it no longer cancels and provides a systematic bias in the weighing measurement of the Kibble balance experiment. The relative size of this \red{effect} can be written very compact, see  equation~(\ref{eq:diaF2}). If the coil and the air gap have the same radius the effect is proportional to the magnetic susceptibility and the ratio of the cross-sectional areas of the coil and the air gap.

Unbeknownst to the scientist and engineers working with Kibble balances, there is also an effect in velocity that arises when a weakly magnetic material is added into the gap. Introducing such a material in the gap changes the magnetic flux density and hence the result that is obtained in the velocity mode. Adding, for example, a diamagnetic coil in the gap reduces the magnetic flux density where the coil is and increases the magnetic flux density in the remainder of the gap.
This is a consequence of the changed reluctance of part of the gap. Where the coil is the magnetic reluctance is larger leading to a smaller amount of flux. However, since the flux through the total air gap remains approximately constant, the flux at the remainder of the gap increases. The increased flux causes a larger induced voltage when the coil is moved through the gap compared to the situation where the coil is non-magnetic. As shown in equation~(\ref{eq:diaV}), the relative change in voltage evaluates to the same expression as for the diamagnetic force. Hence the bias introduced in the weighing mode is cancelled by an equal bias in velocity mode. Thus, the paradox of the diamagnetic force on the Kibble coil is resolved

To prove the existence of the effect of weakly magnetic materials in force and velocity mode we have developed a new technique that we call differential finite element analysis (dFEA). Calculating small forces or field changes caused by the introduction of materials whose susceptibility is of order \SI{1e-5} is impossible. The numerical uncertainties are much larger than the relative effects one desires to calculate. In differential FEA, the susceptibility of the material to be investigated is a parameter and the model is calculated with several different large susceptibilities. Values of $\chi\le \SI{1e-2}{}$ were used, up to a thousand times larger than the susceptibility of the coil in the physical experiment. For differential FEA to work, it is important to keep the same mesh for all calculations. From each calculation result, a null-result that was obtained by setting $\chi$ equal to zero is subtracted. In the end, the quantity of interest is plotted as a function of the used $\chi$ and a smooth function is fitted to the result. The fitted parameters of the function can be used to calculate the effect for small $\chi$ that the physical system has.  For the cases discussed here, both effects scaled linearly with $\chi$ making the scaling simple.

We used differential FEA to calculate the effect that the introduction of a weakly magnetic material has on measurements in force and velocity mode. We find the calculated result in agreement with a simplified analytical model that we have developed in the preceding sections. The relative sizes of the effect are of order \SI{1e-6} and would render Kibble balances useless. The effect has the same magnitude and sign in both modes and will cancel in the combined result. We believe that this is an additional, to date not recognized symmetry of the Kibble balance that allows it to work in the presence of linear magnetic materials. The result of the differential FEA shows that the biases in force and velocity agree within a difference of \SI{1e-3}, limiting the upper bound for the relative bias of the combined measurement to \SI{1e-9}{}.

The paradox of the diamagnetic force in Kibble balances has been solved.  The ongoing discussions in the Kibble balance community are brought to a satisfying end. The reciprocity of Kibble's equation works perfectly in the presence of linear magnetic materials.

\clearpage

\small
\section*{Methods}
\subsection*{Self-inductance $L(z)$ of a coil in a symmetrical yoke}
\label{method1}

Let  $w_a$, $2h_\mathrm{geo}$, and $r_a$ be the geometrically measured width,  height, and  mean radius of the air gap. Neglecting the fringe fields at the end, the air gap has a magnetic reluctance of
$R_\mathrm{ideal}={w_a}/{(4\pi\mu_0 r_ah_\mathrm{geo})}$. The relative correction necessary  to account for the fringe fields scales with  gap's aspect ratio $w_a/(2h_\mathrm{geo})$ \cite{ss20}. The reluctances of the leakage paths at both ends of the gap  are parallel to  $R_a$, lowering the total reluctance of the system. Writing  the reduction factor as $1/\gamma$ with $\gamma>1$, yields $R_a={w_a}/{(4\pi\mu_0 r_ah_\mathrm{geo}\gamma)}$,
which can be interpreted as the reluctance of an ideal air gap of the same dimension, but a magnetic height, $2h_a$, that differs from the geometric one according to $h_\mathrm{a} = \gamma h_\mathrm{geo}$. In our idealized gap, the magnetic flux is purely horizontal. Neglecting the vertical flux, makes the analysis simpler without altering the conclusion.

Now, let's investigate the flux that is produced by a coil at position $z_c$
with $N$ turns carrying a current $I$. The flux $\phi_0$, generated by the magneto motive force of the coil, has to traverse the air gap above and below  the coil. Hence,	
\begin{equation}
\phi_0=\frac{NI}{(R_u+R_l)},~~\mbox{where}~~R_u=\frac{w_a}{2\pi\mu_0 r_a(h_a-z_c)},~~R_l=\frac{w_a}{2\pi\mu_0 r_a(h_a+z_c)},
\label{a1}
\end{equation}
where $R_{u/l}$ denote the magnetic reluctance of the partial gap that is above/below the coil. Using the definition of the self-inductance $L={N\phi_0}/{I}$ and expanding the fractions to second order in $z_c$ yields
\begin{equation}
L=2\pi \mu_0 N^2 r_a\left( \frac{ h_a}{2w_a }-\frac{z_c^2}{2w_a h_a}\right).
\label{eq:method:inductance}
\end{equation}
which appears in the text as equation (\ref{inductance}).
Experimentally, the inductance of the coil can be measured as a function of $z_c$. By fitting  $L=L_0-kz_c^2$ to the data, the magnetic height of the air gap can be determined from $k$ as
\begin{equation}
2h_a=\frac{2\pi\mu_0N^2  r_a  }{w_a k}
\label{21}
\end{equation}

For the magnet employed by the BIPM Kibble balance, $k\approx\SI{550}{\henry \per \meter}$, see \citen{li17}.
Using the reported technical data of that magnet system, $w_a$=13\,mm, $N=1057$ and $r_a=125$\,mm, a magnetic height of $2h_a=\SI{155}{\milli\meter}$ is obtained. A comparison to the measured, geometric height, $2h_\mathrm{geo}=82$\,mm shows that $\gamma=1.89$.

The magnetic height can also be deduced from a finite element analysis (FEA), employing for example equation~(\ref{eq:boundary1}). The value for $2h_a$ obtained by the measurement of $L(z)$ agrees with the one obtained by FEA within a few percent. The difference  is  due to the fact that magnet's top cover was missing in the experiment.

\subsection*{General equation of a moving cylindrical segment with finite $\chi$}

Here, we investigate the effect cause by any weakly magnetic part that is co-moving with the coil. Such  parts are abundant in any Kibble balance experiments and include the coil former, the  supporting frame, optical elements, and fasteners. Without loss of generality, we investigate a cylindrical part with a rectangular cross section identified by the subscript $_i$, its height, width, mean radial location, and magnetic susceptibility are denoted  by $2h_i$, $w_i$, $r_i$, and $\chi_i$. The symmetry axis of the part coincides with the symmetry axis of the magnet.

Starting with equation (\ref{eq:Delta_F_chi}), the diamagnetic force in weighing mode on the segment  is
\begin{equation}
\Delta F_{\chi,i} = \frac{2 \chi_i V_i}{\mu_0}I \partial_z\left( B_{v,i}^2 \alpha \right), \label{eq:Delta_F_chii}
\end{equation}
where $V_i=2\pi r_i(2h_iw_i)$ is the volume of the cylindrical segment and $B_{v,i}$ the magnetic flux density at the segment position without current in the coil. The flux produced by the permanent magnet and the coil are both horizontal, and, hence, the flux density is proportional to $1/r$.
Consequently, at the mean radial position of segment $i$, the flux is $B_{v,i}=({r_c}/{r_i})B_v$, yielding
\begin{equation}
\frac{\Delta F_{\chi,i}}{\Delta F_{\chi}}
= \frac{\chi_i}{\chi} \frac{V_i}{V}\left(\frac{B_{v,i}}{B_v}\right)^2
= \frac{\chi_i}{\chi} \frac{r_c}{r_i}.
\label{fchii}
\end{equation}
Using the expression for the diamagnetic effect on the coil (\ref{eq:diaF2}) together with  (\ref{fchii}), produces a compact expression for the relative diamagnetic force produced by the segment. It is
\begin{eqnarray}
\frac{\Delta F_{\chi,i}}{mg}=-\chi_i \frac{A_i}{A_a} \frac{r_a}{r_i},
\label{fchiirel}
\end{eqnarray}
where $A_i=2h_iw_i$ the sectional area of segment $i$, and $A_a$ the cross sectional area of the air gap defined above.

Most importantly and similarly to equation (\ref{eq:diaF2}), the relative diamagnetic force on segment $i$ is independent of the coil current and the magnetic field $B_v$, and is determined only by the material property ($\chi_i$) and geometrical ratios ($A_i/A_a$ and $r_a/r_i$).

Next, the influence of the weakly magnetic segment $i$ on the measured value of $Bl$ in velocity phase is investigated.
Assuming a  magnet system with perfect up-down symmetry, as shown in figure \ref{fig1}(a), we define the following three surfaces at $r=r_c$: $\mathcal{A}$ ($r\leq r_c$, $z=h_a$) and $\mathcal{B}$ ($r\leq r_c$, $z=-h_a$) present the horizontal surfaces respectively at the upper and lower gap ends. $\mathcal{C}$ ($r\leq r_c$, $z=z_c$) is the coil surface. In perfect symmetry, the magnetic flux $\phi_\mathcal{A}$ penetrating surface $\mathcal{A}$, equals the flux $\phi_\mathcal{B}$ through surface $\mathcal{B}$. An asymmetry can be taken account by introducing a flux difference $\Delta\phi_0$ such that
\begin{equation}
\phi_\mathcal{A}=\phi_\mathcal{B}+\Delta\phi_0,
\label{eq:phi1}
\end{equation}
By using an electrical circuit model following Ohm's law of magnetism as shown in figure \ref{fig1}(a), $\phi_\mathcal{A}$ and $\phi_\mathcal{B}$ are determined as
\begin{equation}
\phi_\mathcal{A}=\frac{E_1}{(R_m+R_a)}\;\;\mbox{and}\;\;\phi_\mathcal{B}=\frac{E_2}{(R_m+R_a)}.
\end{equation}
A magnetic segment co-moving with the coil, does not contribute to the air gap reluctance $R_a$, and, hence, $R_a$ does not depend on the vertical position of the segment. Since $\phi_\mathcal{A}$, $\phi_\mathcal{B}$ are constant for a given magnet system, the flux difference $\Delta\phi_0$ must also be independent of the coil position $z_c$.

That flux that goes through surface $\mathcal{A}$ will then either go through the coil or through the part of the air gap that is above the coil $\phi_\mathcal{U}$, $\phi_\mathcal{A}=\phi_\mathcal{C}+\phi_\mathcal{U}$. Similarly, all the flux penetrating $\mathcal{B}$ flows through the part of the air gap that is below the air gap or through the coil,
$\phi_\mathcal{B}=-\phi_\mathcal{C}+\phi_\mathcal{L}$. The negative sign before the coil flux indicated the direction of the flux relative to the normal vector of the coil. It is reverse for the flux $\phi_\mathcal{B}$.

The fluxes $\phi_\mathcal{U}$ and $\phi_\mathcal{L}$ can be written as a product of the surface area and the magnetic field at the radius $r_i$ under the assumption that the field is mostly independent of $z$. Hence,
\begin{eqnarray}
\label{phiul}
\begin{cases}
\phi_\mathcal{U}(\chi_i=0)=2\pi r_i (h_a-z_c)\,B_{0,i}, \\
\phi_\mathcal{L}(\chi_i=0)=2\pi r_i(h_a+z_c)\,B_{0,i}.
\end{cases}
\end{eqnarray}
the flux through the coil can now be obtained as
\begin{equation}
\phi_\mathcal{C}(\chi_i=0)=2\pi r_iB_{0,i}z_c+\frac{\Delta\phi_0}{2}.
\end{equation}
To evaluate the induced voltage only the component that depends on $z$ is relevant and we obtain
\begin{equation}
U(\chi_i=0)=2\pi r_iNB_{0,i}v_z.
\label{p5}
\end{equation}
Taking again advantage of the $1/r$ dependence of the magnetic flux density,  $B_{0,i}=r_c/r_i B_{0}$. For $\chi=0$, the voltage-velocity ratio is the $2\pi r_cNB_{0}=B_0l$ in agreement with the conventional theory of the Kibble balance.

For $\chi_i\neq0$, the magnetic field distribution along the vertical direction at $r_i$ is given by discrete two values with sharp steps between them. In the coil region (from $z_i-h_i$ to $z_i+h_i$), the magnetic flux density is $B_{c,i}$ and the magnetic flux density of the rest air gap is $B_{\chi,i}$. In this case $\phi_U$ and $\phi_L$ are written as
\begin{eqnarray}
\label{p4}
\begin{cases}
\phi_\mathcal{U}(\chi_i)=2\pi r_i\left[B_{c,i}\eta_i 2h_i+B_{\chi,i}\left(h_a-z_c-\eta 2h_i\right)\right],\\
\phi_\mathcal{L}(\chi_i)=2\pi r_i\left\{B_{c,i} (1-\eta_i)2h_i+B_{\chi,i}\left[h_a+z_c-(1-\eta_i) 2h_i\right]\right\},
\end{cases}
\end{eqnarray}
where $\eta_i$ defines the height fraction of the segment that is above the coil $z_c$. For example, when segment $i$ is fully above the coil, $\eta_i=1$. If segment $i$ is coincident with the coil (like the coil itself), then $\eta_i=0.5$. Note since the segment is co-moving with the coil, $\eta_i$ does not change with $z_c$. Through equation (\ref{p4}) and known magnetic flux relations, the magnetic flux through the coil is solved as
\begin{eqnarray}
\phi_\mathcal{C}(\chi_i)&=&2\pi r_i\left[B_{\chi,i}\left(z_c+(2\eta_i-1)h_i\right)-B_{c,i}(2\eta_i-1)h_i\right] +\frac{\Delta\phi_0}{2}.
\end{eqnarray}
Dismissing all factors that are independent of $z_c$, the induced voltages simplifies to
\begin{equation}
U(\chi_i)=\frac{N\partial\phi_C(\chi_i)}{\partial t}=2\pi r_iNB_{\chi,i}v_z.
\label{p6}
\end{equation}
Comparing equation (\ref{p5}) to equation (\ref{p6}), the relative change in the induced voltage cause by segment $i$ is
\begin{equation}
\frac{\Delta U_{\chi,i}}{U}=\frac{B_{\chi,i}}{B_{0,i}}-1.
\end{equation}
This result is the equivalent expression as given in equation (\ref{eq:Delta_U}). The result of $B_\chi/B_0-1$ remains valid and the relative change in the induced voltage is
\begin{equation}
\frac{\Delta U_{\chi,i}}{U}=\frac{B_{\chi,i}}{B_{0,i}}-1=-\chi_i \frac{A_i}{A_a} \frac{r_a}{r_i}.
\label{vf}
\end{equation}

The relative effects in force and velocity mode caused by the introduction of a cylindrical segment $i$ with finite susceptibility are identical, compare equation (\ref{fchiirel}) to equation (\ref{vf}).

The relative changes of the induced voltage and the extraneous force produced by a single segment $i$ depend on the  ratios $A_i/A_a$ and $r_i/r_a$.  Hence, the scaling can be checked by comparing the relative effects produced by two different segments $i$ and $j$ with the same magnetic susceptibility. The ratio of the relative effects must scale like $(A_i/r_i)/(A_j/r_j)$. The calculation was performed for both segments for both modes, velocity mode and force mode. Again, differential finite element analysis as described in the main text was used to interpolate the effects to $\chi_j=\chi_i=\SI{-1e-5}{}$ by using calculations that used susceptibilities ranging from -0.01 to 0.01.

As segment $i$, we use the coil which has been already shown in the main text and, for convenience, we reiterate the numbers here. The coils has a cross-sectional area of $A_i=\SI{200}{\milli \meter^2}$ and a mean radius of $r_i=\SI{125}{\milli\meter}$. It produces a relative effect of $\SI{9.50e-7}{}$.

For the segment $j$ we chose an area of $A_j=\SI{50}{\milli \meter^2}$. It is located at $r_j=\SI{122.5}{\milli\meter}$. To also check if the calculation works for objects that are offset from the vertical coordinates of the coil, we placed segment $j$ \SI{17.5}{\milli\meter} above the coil.  The relative effect for segment $j$ is the same for force and velocity mode and it is $\SI{2.43e-7}{}$.

The ratio of the two relative effects is 3.92 and the ratio of the corresponding geometrical factors $(A_i r_j) / (A_j r_i)=3.92$. The agreement between the geometrical ratios and the calculated effect validate equations (\ref{fchiirel}) and (\ref{vf}).

\section*{Acknowledgments}
During the last two years, many experts in the field (BIPM, NIST, NRC, NIM) have been involved in this study. We would like to thank all these colleagues for their valuable discussions with us, especially Michael Stock (BIPM) and Carlos Sanchez (NRC). Their advice was helpful to understand the effect and to finally solve the paradox out. We also would like to thank Jon Pratt (NIST) for the werb review and constructive suggestions to improve the manuscript quality.

\section*{Author contributions statement}

S. Li and S. Schlamminger developed the analytical theory. Major discussions have been ongoing among all authors for more than two years. Numerical simulations were preformed by S. Li and R. Marangoni. The manuscript was written by S. Li and S. Schlamminger and all authors reviewed the manuscript.

\section*{Competing interests}
The authors declare no competing interests.

\clearpage
\section*{Supplementary Information}
\section*{The magnetic field produced by a coil in free space and inside a yoke}

\begin{figure}[tp!]
	\centering
	\includegraphics[width=0.6\textwidth]{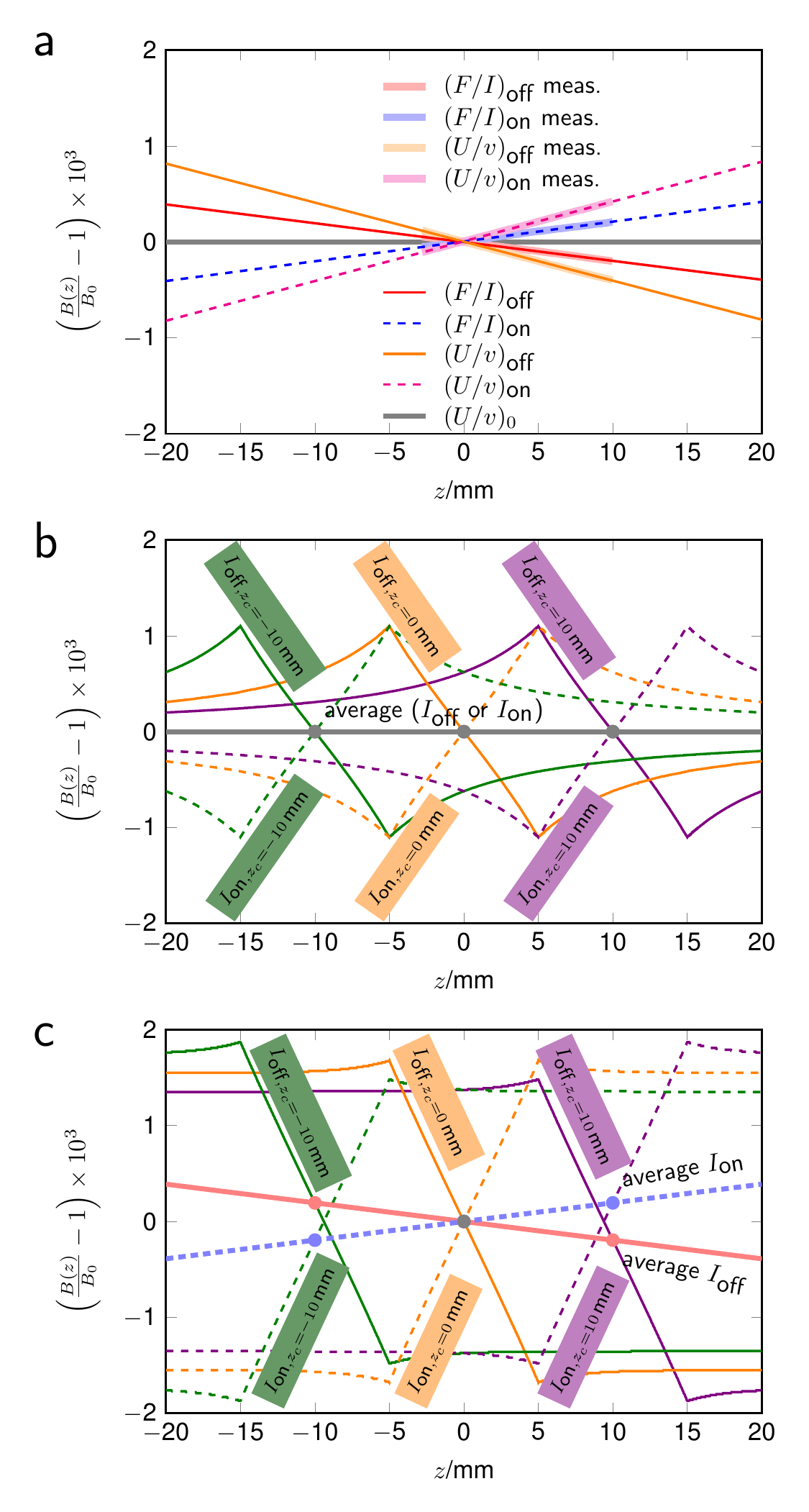}
	\caption{\small(a) The magnetic field change due to the coil current. The middle thick curves are obtained by experimental measurement in the BIPM system\cite{li17} and the thin lines are linear extrapolations from -20\,mm to 20\,mm. (b) presents the relative $B$ field distribution when the yoke permeability is set to $\mu_0$ and (c) shows the magnetic field change with soft normal yoke. Note the reference field, $B_0=0.45$\,T, is used for both cases (b) and (c).}
	\label{fig:s1}
\end{figure}

In this section, we compare the magnetic field of a current-carrying coil in free space and in a Kibble balance magnet. Figure \ref{fig:s1}(a) shows the effect that the current in the coil has on the measured $Bl$ in force and velocity mode.
We present a summary of the finding that have been described in~\cite{li17}. Plotted on the vertical axis is the difference of $Bl$ measured in different scenarios from a velocity phase measurement $(U/v)_0=B_0l$ where the coil is moving through the magnetic field without any current. The two profiles  $(F/I)_{\mbox{off}}$ and $(F/I)_{\mbox{on}}$ are used in most Kibble balances for weighing -- usually the two measurements in weighing phase, mass-on and mass-off, are carried out with equal and opposite currents, $I_{\mbox{off}}=-I_{\mbox{on}}$. The profiles  $(U/v)_{\mbox{off}}$ and $(U/v)_{\mbox{on}}$ are profiles determined in the velocity phase, respectively with plus and minus currents. The bifilar coil used in the BIPM balance allows current to be present during the velocity mode. For conventional two-mode Kibble balances, $(U/v)_0$ is used, while $(U/v)_{\mbox{off}}$ and $(U/v)_{\mbox{on}}$ are for one-mode measurement schemes, e.g. \cite{BIPM,UME}.

The data in figure \ref{fig:s1}(a) shows that the change in $Bl$ is a linear in coil position $z$, as well as in coil current $I$ in the weighing measurement. As discussed in the article,  the field gradient $\partial_z B$  produces a diamagnetic force bias that can not removed by mass-on and mass-off measurements.
The interesting conundrum is, can the magnet field gradient that is produced by the coil itself produce a force on itself. 

To clarify the paradox we study two scenarios: 1) the coil in free space and 2) the coil inside an air gap surrounded by iron yokes as it would in a Kibble balance . Figure \ref{fig:s1}(b) and (c) show  the magnetic flux density distribution along $z$ for different weighing positions and both current directions. The results shown in the figure are obtained by finite element analysis based on the BIPM magnet system \cite{BIPMmag2017} with an  average field in the air gap of  $B_0=0.45$\,T. To obtain the free-space calculation, the yoke relative permeability is simply set to one. Using the same geometry and mesh, results in a trustworthy comparison of both scenarios.

Figure \ref{fig:s1}(b)  clearly shows that the magnetic field distribution is independent of the vertical position for the coil in free space. The size and shapes of the green, orange and violet lines are identical only horizontally displaced by 10 mm from each other. Hence, energetically it does not matter where the coil sits. There is no minimum and hence no force on the coil. Although magnetic gradients exist on the curve $B(z)$ of a free space system, any force related to these gradients is internal (the coil gets compressed), which cannot be seen by the weighing unit (balance or mass compactor). 

For the coil in  he yoke, shown in figure \ref{fig:s1}(c), the situation is different. The additional fields are only symmetric for $z=0$\,mm. At all other positions ($z\neq 0$\,mm) the symmetry is broken. To one side of the coil the magnetic flux density is higher to the other it is lower.  As was shown above, it is not the local gradient of the field that provides  information on the force.  The volume average  over the coil must be considered. The volume average, shown as the blue and red line in the figure, is no longer independent of $z$ as it was in the free space scenario.

The external force on the coil depends on the average field. The experimental observation, shown in figure~\ref{fig:s1}(a) agrees well with the FEA calculation presented in~\ref{fig:s1}(c).

The comparison of average fields produced by a current-carrying coil in free space and in  a yoke shows that in the latter the broken symmetry  will lead to an energy redistribution and  hence a magnetic field change. A physical picture can be provided as follows: In the yoke $B$ changes because the coil current magnetizes the yoke, and the magnetized yoke produces an additional magnetic field at the coil position.

The diamagnetic force arises from $F_\chi =\frac{\chi V}{\mu_0}B \partial B_z$, where the last factor is the derivative of the volume average field with respect to the coil position. As shown in figure~\ref{fig:s1}(c), the slopes of  the red and blue lines and, hence the gradient $\partial B_z$ is constant within reasonable ranges of $z$. Accordingly, the diamagnetic force is constant as a function of $z$, but changes direction when the current is reversed. Because of the latter fact, the diamagnetic force does not cancel in the Kibble balance experiment and can lead to a systematic bias.

In conclusion, the statement that `a current-carrying coil can not produce a measurable force on itself' holds for free space system, but not necessarily for  a coil inside a yoke, unless the yoke moves with the coil. We note that this finding does not contradict Newton's third law as there is an equal and opposite force on the yoke.

\section*{Differential FEA (dFEA) -- A method to calculate small effects}

The diamagnetic effects described in the main article are very small. In force mode,  an additional bias of \SI{10}{\micro \newton} on top of \SI{10}{\newton} is produced. It would be impossible with finite element analysis to detect, let alone to calculate with any uncertainty, this additional bias. We first developed the analytical equations of the diamagnetic effect in force and velocity modes. However, we wanted to verify our analytic results with numerical calculations. In searching for ways to do this, we invented differential FEA (dFEA). With differential FEA, it was possible to calculate the size of the effect with a relative uncertainty of about $0.1$\%, corresponding to f \SI{10}{\nano \newton}. We could further show that the diamagnetic effect produces the same bias in force and velocity mode within that uncertainty. 

Conventional finite element analysis, FEA, is a powerful technique to solve a variety of engineering problems that would be difficult or impossible to solve analytically. 
The basic idea of FEA is to divide the domain under investigation into small elements, and then to approximate the solution of the problem by a linear combination of calculations in each element. 
In FEA simulations, small errors are inevitable. Either discretization or numerical errors cause these problems. The computer simulates a discrete system, while the real system in nature is continuous. Numerical errors can occur, for example, by rounding when two large numbers are subtracted or by ill-conditioned matrices. 
In our experience, problems in magnetostatic can be calculated with relative errors of a few parts in $10^{3}$ on a standard PC in reasonable time.
The effects discussed in our work are of the order $1\times10^{-6}$. Hence, they are three orders of magnitude smaller than what we consider reliable results obtainable by FEA.

With dFEA,  small effects and their uncertainty can be calculated using commercial FEA packages without exponentially extending the computation time. The technique takes about five times longer than a single calculation because, as the reader shall see below, the same model has to be calculated about five times with different $\chi$.
The idea is simple. As mentioned above, the uncertainties of the calculations are 1000 times larger than the effect, so if one can increase the effect by a factor of 1000, the effect could be detected.
In our case, where the effect scales with the susceptibility, one only needs to multiply the susceptibility of the part that is investigated by a factor of 1000 in question by a factor of 1000 to achieve that amplification. So, if $\chi_{\mathrm{nom}}$ is the susceptibility of the part in question, here the coil, $\chi_{\mathrm{exag}}\approx 1000 \times \chi_{\mathrm{nom}}$ is used in the finite element calculation.
It is safe to change the susceptibility because it occurs as a linear parameter, and the underlying physics does not change.
In the end, all one has to do is scale the measured effect, say the force, by $\chi_{\mathrm{nom}}/\chi_{\mathrm{exag}}$ to obtain the size of the previously immeasurable effect. 

\begin{figure}[tp!]
	\centering
	\includegraphics[width=0.6\textwidth]{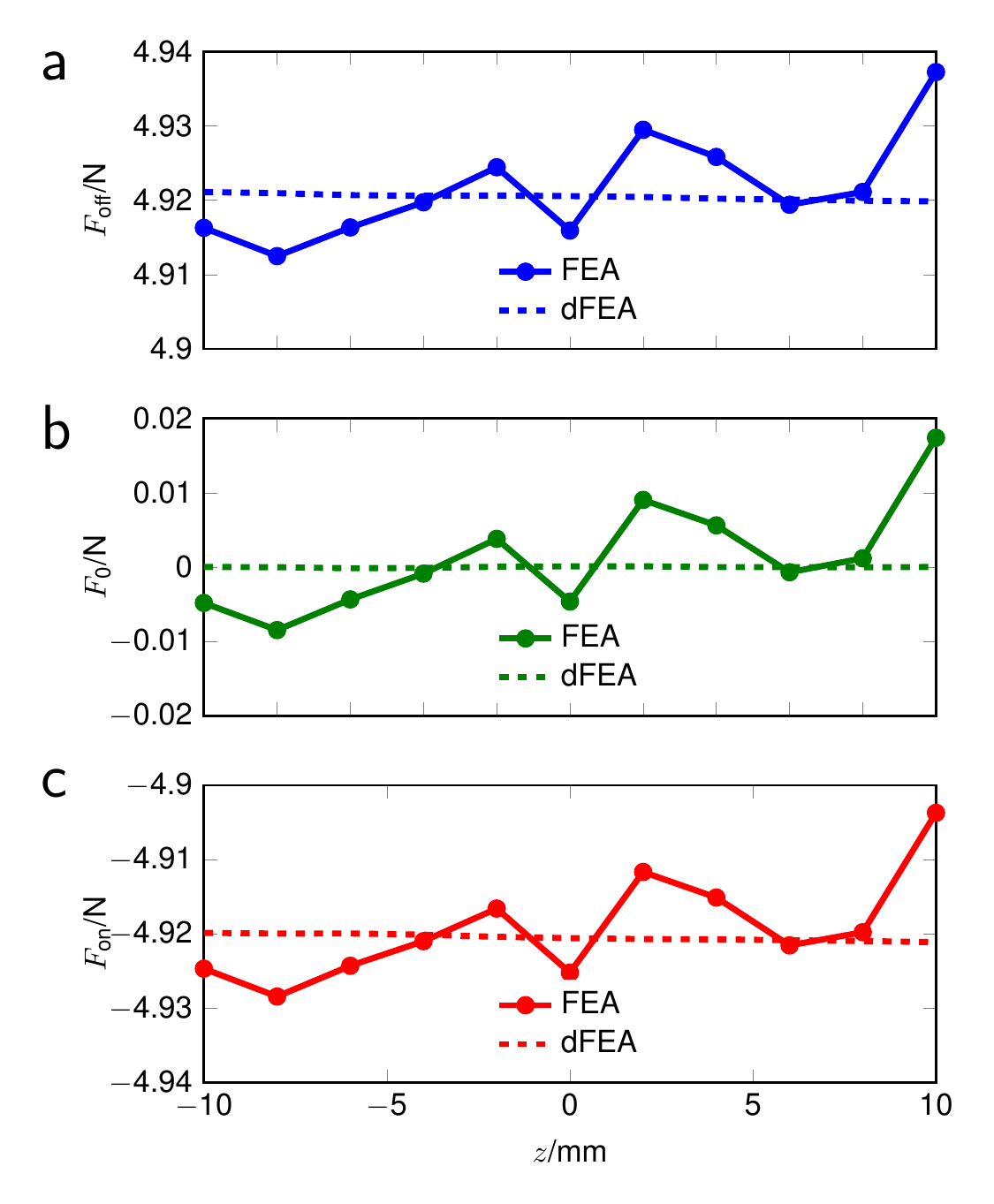}
	\caption{\small Simulation results of the magnetic force of a coil in a magnetic field  with different currents: (a) $I=I_\text{off}$, (b) $I=0$ and (c) $I=I_\text{on}$. Note that this example is the case with $\chi=-0.01$. 
	The solid lines are obtained from FEA calculations, while in (a) and (c) the dashed lines are differential signals, $F_\text{off}-F_0$ and $F_\text{on}-F_0$. 
	The dashed line in (b) is half the difference of the two forces, $(F_\text{off}-F_\text{on})/2$. In order to plot it in the same scale the mean of \SI{4.92}{\newton} has been subtracted. 
	}
	\label{fig:s2}
\end{figure}

A successful implementation of dFEA relies on three good practices:
\begin{enumerate}
\item It is not sufficient to  perform one calculation with $\chi_{\mathrm{exag}}$  and scale the result by $\chi_{\mathrm{nom}}/\chi_{\mathrm{exag}}$. Usually there are small calculation biases that would be scaled and falsify the result. See, for example, the  middle graph of figure \ref{fig:s2} , where the force calculated with zero current is not exactly zero. The best practice is to calculate the desired effect for several $\chi$ values and then interpolate the size of the effect to the nominal $\chi$. For example, we would like to calculate the diamagnetic effect for $\chi_{\mathrm{nom}}=\SI{1e-5}{}$. We use five value for $\chi_{\mathrm{exag}}$, namely 
-0.01, -0.005, 0, 0.005, 0.01. We then plot the desired function, for example the force $F$, as a function of $\chi_{\mathrm{exag}}$, subtracting $F(\chi_{\mathrm{exag}} = 0)$. This function is linear in $\chi_{\mathrm{exag}}$  and the value at $\chi_{\mathrm{nom}}$ can be obtained from a linear fit to the data.

\item It is of utmost importance to keep all other parameters constant when using dFEA. If additional parameters were changed for the calculation with different $\chi_{\mathrm{exag}}$, they would cause a change in the calculation result that would then, falsely, be attributed to being driven by the susceptibility. This point applies, especially to the meshing. The domain should be meshed only once before the various  $\chi_{\mathrm{exag}}$ are assigned to the components.

\item The third practice applies to calculations where current is involved. In this case, the best results are obtained by calculating the system twice, with and without current. The difference between the two results gives the desired effect. Again, other than the current, nothing can change between the calculations, most importantly, not the meshing.
\end{enumerate}

Figure~\ref{fig:s2} illustrates the importance of the three practices mentioned above. The figure shows the  FEA results of the force on the coil at eleven $z$ positions for three different currents, $I_\text{off}$, 0, and  $I_\text{on}$. The forces, as a function of $z$, show the same trend for all three currents. The trend is caused by numerical biases in the FEA software and affects all three curves in the same, nonphysical, fashion.  By subtracting the data obtained with $I=0$, the numerical biases can be subtracted out. The curves labeled dFEA show relative results.

The curves obtained with dFEA are not only smooth, but they are also physical. They represent the coil inductance force \cite{li17}, a linear force curve over $z$. A physical result can be obtained from the original noisy result. The standard deviation of the data calculated with dFEA is 100 times smaller than standard FEA data.

In summary, dFEA allows one to calculate a relative diamagnetic force of $1\times10^{-6}$ with meaningful uncertainties. The method relies on FEA calculations with several exaggerated susceptibilities, the largest one about 1000 times the size of the nominal susceptibility. 

\end{document}